# A Reputation-based Approach using Consortium Blockchain for Cyber Threat Intelligence Sharing

ZHANG Xiaohui[①]     MIAO Xianghua*[①②]

[①](*Faculty of Information Engineering and Automation, Kunming University of Science and Technology, Kunming, China*)

[②](*Computer Technology Application Key Laboratory of Yunnan Province, Kunming, China*)

**Abstract**. The CTI (Cyber Threat Intelligence) sharing and exchange is an effective method to improve the responsiveness of the protection party. Blockchain technology enables sharing collaboration consortium to conduct a trusted CTI sharing and exchange without a trusted centralized institution. However, the distributed connectivity of the blockchain-based CTI sharing model proposed before exposes the systems into byzantine attacks, the compromised members of partner organizations will further decrease the accuracy and trust level of CTI by generating false reporting. To address the unbalance issues of performance in speed, scalability and security, this paper proposes a new blockchain-based CTI model, which combines consortium blockchain and distributed reputation management systems to achieve automated analysis and response of tactical threat intelligence. In addition, the novel consensus algorithm of consortium blockchain that is fit for CTI sharing and exchange introduced in this paper. The new consensus algorithm is called "Proof-of Reputation" (PoR) consensus, which meets the requirements of transaction rate and makes the consensus in a creditable network environment through constructing a reputation model. Finally, the effectiveness and security performance of the proposed model and consensus algorithm is verified by experiments.

## 1 INTRODUCTION

In order to mitigate the danger of increasingly complex attack methods or threats such as advanced persistent threats (APTs) and zero-day vulnerabilities that brought about by the development of information technology in recent years, organizations need to be supported by more effectiveness and responsiveness defense methods. As the proactive approach, CTI (Cyber Threat Intelligence) is a collection of information that can cause potential harm and direct harm to organizations and institutions [1]. The typical application of CTI is shown in Fig. 1. CTI has become an important weapon in the arsenal of cyber defenders to address the information asymmetry of issues that happened in offensive and defensive sides, taking advantage of the value behind the CTI such as evaluate and simulate malicious behavior in networks is the key measure to mitigate increasing cyber-attacks.

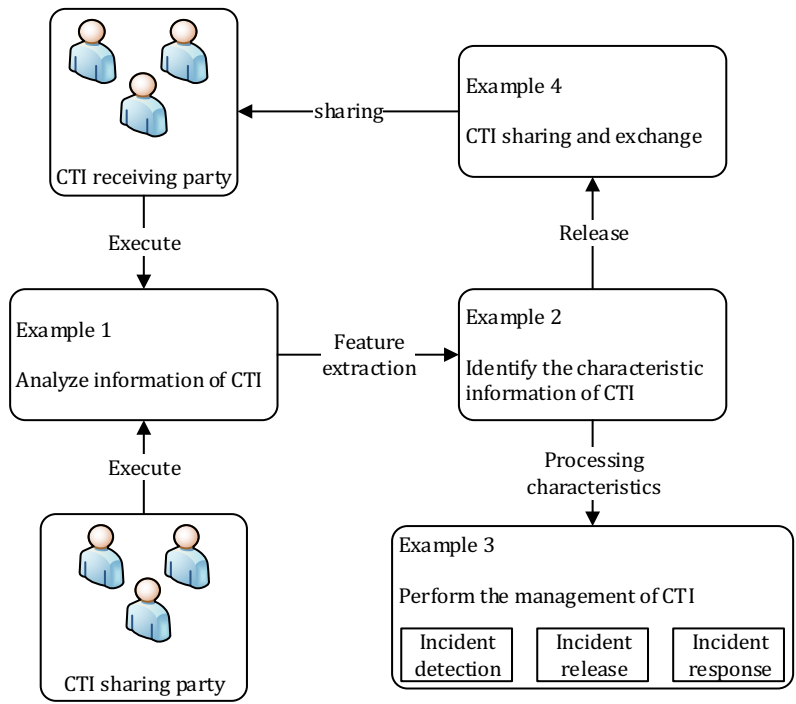

**Fig 1. The application of cybersecurity threat intelligence**

  The CTI sharing and exchange in a co-operatively approach, promises to be the most effective method to maximize the benefit of CTI through improve the issue of information islands, which means the CTI generated from partner organizations can aid cybersecurity policymakers in making decisions. In order to meets the needs of CTI sharing, the stakeholders has formulated a series of standards for the exchange of threat intelligence, such as STIX, IODEF, OpenIoC [2]. The typical structure of CTI sharing system is shown in Fig. 2. The core idea behind threat intelligence sharing is to create situation awareness among stakeholders through sharing information about the newest threats and vulnerabilities and to swiftly implement the remedies [3]. However, the result of a survey conducted in 2014 shows that slow and manual sharing processes impede full CTI exchange participation [4]. Therefore, automated process to defend against the attacks need to be realized in CTI sharing platform, which is helpful to take efficient protect action in cyber threats. For example, there have been large-scale WannaCry virus in education, medical and other industries in recent years [5], if these threat intelligences can be timely released, then most organizations will be able to avoid intrusion, which means that automating the sharing and exchange processes can extremely increase the effectiveness of CTI.

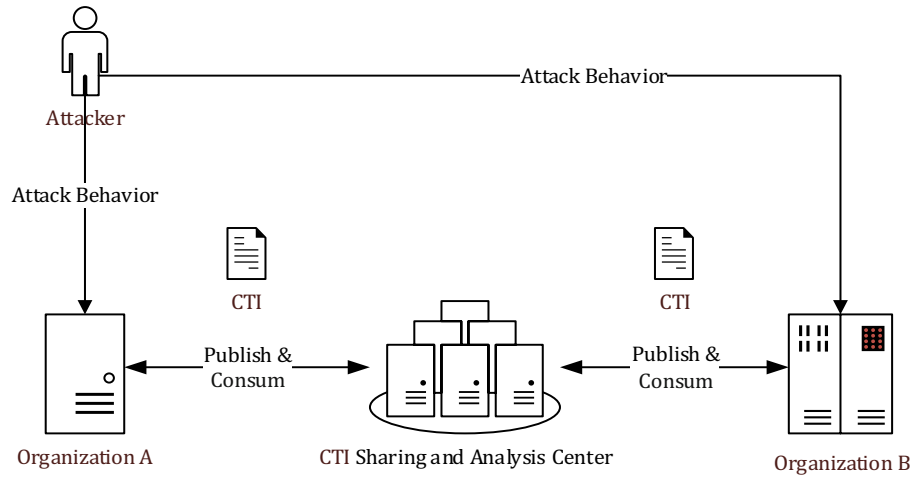

**Fig 2. The typical structure of CTI sharing system**

According to the different applications of CTI, it can be categorized as tactical threat intelligence, operational threat intelligence, strategic threat intelligence, and technical threat intelligence [6]. Tactical threat intelligence is consumed by incident responders to ensure that their defenses and investigation are prepared for current tactics [7]. Consequently, the key to achieve CTI sharing automation is to be accurately received and processed about a tactical threat intelligence in a quickly way.

The inappropriate CTI sharing may lead to the disclosure of critical and sensitive intelligence data that included in CTI, which can affect the enthusiasm of enterprises to participate in CTI exchange [8]. Hence, there is still a contradiction between automated sharing and the privacy protection requirement in CTI sharing platform. The blockchain-based CTI sharing model has brought hope to solve the above contradiction [9]. As a novel framework, blockchain technology which uses the account anonymity, tamper-free mechanism, and encryption function enable sharing participants to conduct a trusted CTI sharing and exchange without a centralized institution [10]. However, the distributed connectivity of the blockchain-based CTI sharing model exposes the systems into various challenges.

On the one hand, in a distributed environment, CTI sharing platform is vulnerable to 'false reporting' issues, which is caused by federation members maliciously reporting cyber-attack intelligence [11]. Byzantine behaviors that happened in the blockchain system may decrease the trust of each other among the members of CTI sharing collaboration consortium [12]. On the other hand, the high throughput is very important to achieve interoperability in CTI sharing and exchange [13], but many studies implemented blockchain solutions through the use of flawed consensus algorithm to exchange data, the limitations of performance and scalability are still exist in these consensus algorithms [14].

Therefore, a new model which combines consortium blockchain and distributed reputation management systems to achieve automated sharing of tactical threat intelligence is presented. The main contributions of our work are summarized as follows:

(1) The model proposed in this paper develops a mechanism to automate sharing of CTI by built a decentralized collaboration consortium, and further addresses the security requirements such as protection against byzantine behaviors.

(2) This paper a proposes reputation model and "Proof-of-Reputation" (PoR) consensus algorithm that integrates the appropriate reputation computing model to meet the requirements of transaction rate and accuracy in CTI sharing, which is together with the distributed reputation management system, to reduce the impact of the byzantine behaviors that happened in the blockchain-based CTI sharing collaboration consortium.

## 2 RELATED WORKS

### 2.1 Consensus Algorithm in Blockchain

Blockchain is a distributed ledger behind bitcoin, which was founded by Satoshi Nakamoto in 2008[15]. As the foundation and core technologies of the blockchain system, the consensus algorithm is critical for the security and performance of the blockchain [16]. Although the public blockchain consensus algorithm such as PoS and DPoS solves the problem that computing power consumption in PoW, there are still problems such as low efficiency, high resource cost, and small throughput [17]. Consortium blockchain consensus algorithm such as Raft [18] can achieve high throughput, but it can only be suitable for non-byzantine environments that only honest nodes in the network [19]. Table 1 presents a comparison of different types of consensus algorithm in the blockchain.

Table 1. Comparisons between two types of consensus algorithm in blockchain

| Criteria | Crash Fault Tolerance | Byzantine Fault Tolerance |
| --- | --- | --- |
| The basis of agreement | Mostly are Voting-based | Mostly are Proof-based |
| Decentralization | Low | Mostly high |
| The way of nodes management | Join network need to be authorized | Join network freely |
| Award | Mostly no | Yes |
| Security | Mostly lower | Mostly higher |
| Speed | Fast | Low |

Many researchers want to use the Byzantine Fault Tolerance (BFT) mechanism to optimize the security performance of consortium blockchain consensus algorithm. The traffic complexity and scalability of Practical Byzantine Fault Tolerance (PBFT) algorithm is the main reason to limits the application of which [20]. Chen et al. proposed a Raft blockchain consensus algorithm based on credit model (Craft), which can be used in a byzantine network environment in 2018[21], experimental results show that the CRaft algorithm has better performance than PBFT, but it still exists 17.89% false positive rate of byzantine nodes. The new consensus algorithm, Proof-of-Trust (PoT), that suitable for crowdsourcing services was proposed in 2019[22], the PoT can provide a feasible accountability method for the application of online services used blockchain technology by selecting the validator of the transaction based on the trust value of the service participants. In 2020, Wang et al. developed the Beh-Raft algorithm [23], which combines the Proof-of-Behavior algorithm (PoB) and Raft algorithm, ensure that only honest nodes can become the leader in the network to reduce the impact of byzantine nodes.

Many related algorithms still cannot meet the scenarios of CTI sharing in an efficient way and require pre-designed malicious behavior models. However, in an untrust network environment, the imbalance in the number of normal and malicious nodes makes it difficult to construct an accurate classifier. So, it is necessary to develop a new consensus algorithm to achieve better performance trade-offs in efficiency and security.

### 2.2 Blockchain-Based CTI Sharing Model

Blockchain technology can enable sharing partner organizations to conduct a trusted CTI sharing and exchange without a trusted centralized institution. There are many studies of the blockchain-base CTI sharing approach carried by researchers, which provide a basis and reference for this paper.

A blockchain-based CTI sharing framework, iShare, proposed in 2018[24], where members participating in the framework can only share the experience of network security protection, iShare uses game theory to analyze malicious behaviors within the framework. Huang et al. published a blockchain-based CTI exchange model in 2019[25], which uses the one-way encryption function to protect the privacy information of participating organizations and analyze the complete network attack chain. In response to the trust and privacy protection issues in CTI sharing, Wang et al. used the technology of channel and membership manager in consortium blockchain to enable trusted participants to disseminate highly sensitive data privately [26].

Collaborative Intrusion Detection Systems (CIDN) [27] is one of the specific applications of tactical threat intelligence. To eliminate insider attacks such as random poisoning attacks and special on-off attacks, improve the accuracy and effectiveness of threat intelligence in CIDN, a threat intelligence aggregation algorithm based on Bayesian decision is proposed by Carol J. Fung et al.[28], which reduces the risk cost of wrong decisions effectively. Wenjuan Li et al. uses blockchain technology to enhance the robustness of the threat intelligence sharing system and protect against insider attacks that may occur during the intelligence aggregation process in CIDN[29]. Chandralekha Yenugunti et al. published a new consensus algorithm based on the trust value of nodes [30], by using the IDS component of each node in blockchain to verify the traffic log and evaluate the credibility of the threat intelligence received from others.

The existing approaches suffer from main issues that impossible to determine whether the generated CTI has been tampered with due to malicious attacks. Our work on CTI sharing model is motivated by the above works, and for incentivizing federation members via a distributed reputation management system [31].

## 3 THE PROPOSED ARCHITECTURE

According to the sources of CTI, threat intelligence can be divided into internal and external [32]. Internal threat intelligence is generally produced from the security devices and system event logs of organizations; external threat intelligence includes commercial threat intelligence sold by the cybersecurity service provider, and open-source threat intelligence that is shared on public network platforms.

The architecture of CTI sharing and exchange using consortium blockchain is shown in Fig. 3. The original CTI are obtained by CTI partner organizations from external, they can also be triggered when an abnormal state is found out by the internal cybersecurity system. Each member in CTI sharing collaboration consortium which composed of proposal generation component, intelligence decision component, and analysis component.

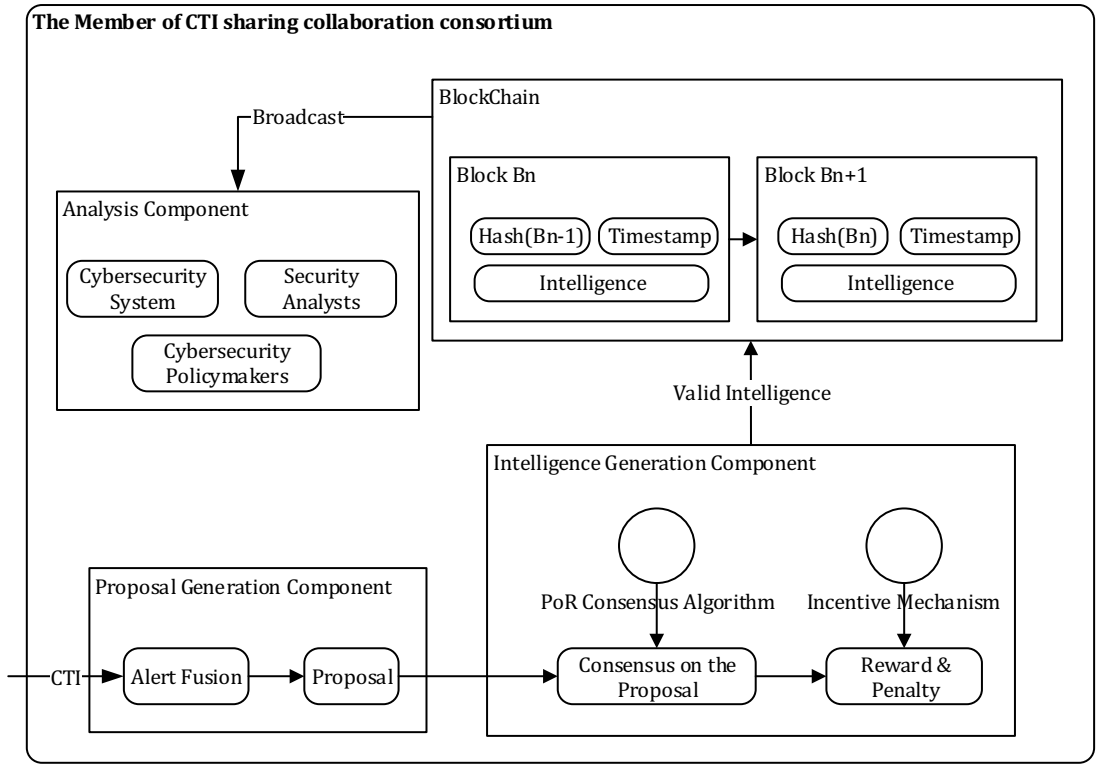

**Fig 3 The architecture of our approach**

The proposal generation component work for generate proposals that transformed from the original CTI for the CTI consortium network, which is used to protect the private information in CTI. Compared to the original CTI, the proposal only includes the key information such as attack characteristics or IP address of the attacker. Proposal results will submit to the intelligence generation component for further processing.

The intelligence generation component realizes the consensus and transmission of proposal among CTI consortium networks and stores the results in blockchain to ensure its immutability and reliability by a innovative consensus algorithm that is fit for the CTI sharing and exchange called "Proof-of Reputation" (PoR). This algorithm makes the consensus of the proposal in a creditable network environment through constructing a reputation model. The content of the PoR consensus algorithm and reputation model will be elaborated in Chapter 4

The receive and respond component represents the cybersecurity policymakers such as the security operations center and security analysts. The intelligence received from the intelligence generation component will be processed further by the member of CTI sharing collaboration consortium to treatment decisions that provide information support or automated take response.

## 4 THE PROOF-OF-REPUTATION CONSENSUS ALGORITHM
### 4.1 Basic Definitions

To address the problem that the consensus algorithm of consortium blockchain cannot meet the transmission requirements of CTI sharing and exchange or only be used in the non-byzantine environment, this paper proposes a consensus algorithm based on reputation model - "Proof-of Reputation" (PoR).

**Definition 1: The State of nodes in PoR**. These nodes of PoR are in one of the following

four states, leader, candidate, follower, and supervisor, the description of four are described in Table 2.

Table 2. The description of four states in PoR

| State | Responsibilities | Description |
| --- | --- | --- |
| Leader | lead the consensus process until its term has expired or to be unfaithful node. | Only one exist in network |
| Follower | Response the request from leader or candidate. | / |
| Candidate | The intermediate state of follower and leader. | Not long-lived in the network. |
| Supervisor | Evaluates the result of proposal that based on all follower's detection. | Part-time by the follower |

The supervisor who is part-time by the follower evaluates the result of the proposal that based on all follower's detection in the network does not change the feature of decentralized in blockchain fundamentally. The conversion relationship of the four states is shown in Fig. 4.

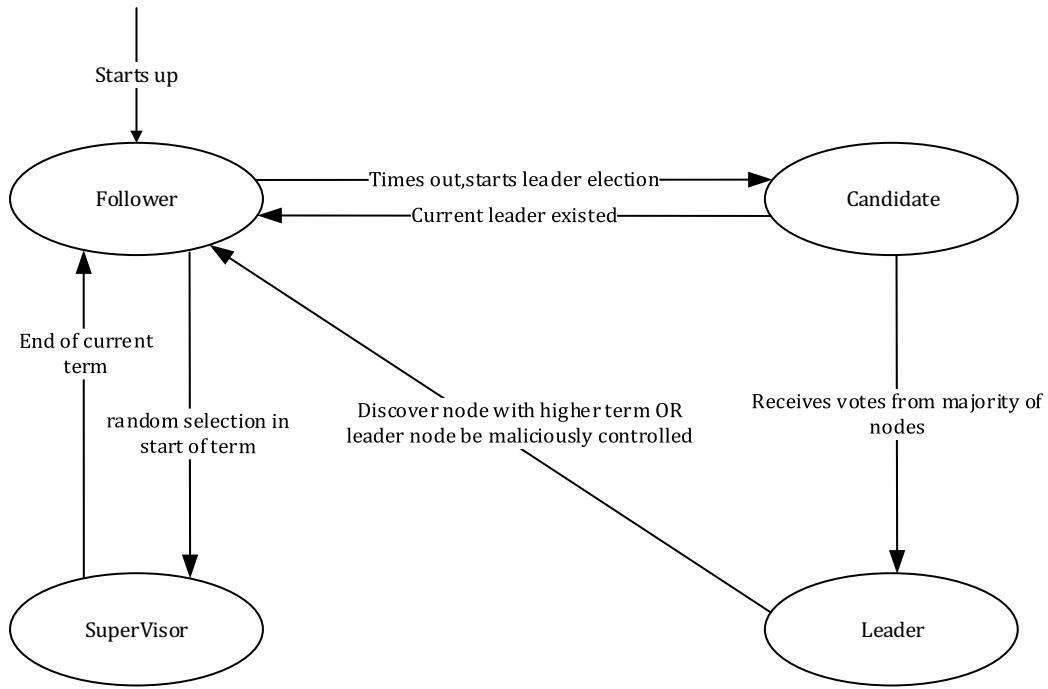

Fig 4 The conversion relationship in PoR

**Definition 2: The type of nodes in PoR.** The type of nodes in PoR are divided into two categories, faithful node, and unfaithful node. A faithful node that exhibits the behavior that making the right decision on the proposal. An unfaithful node means a node that make the wrong decision on the proposal because lack of enough experience or generate false reporting in order to decrease the accuracy of the proposal due to undier malicious control.

**Definition 3: Reputation score.** The reputation score of the node means the probability of peers to provide reliable information, which is used to determine the type of nodes in PoR. Reputation score expressed by $R_i \in [1,100]$. Initial reputation score $R_{init}$ is the constant that indicated the trust level of the new node. A node's reputation score will be calculated according to the behavior and performance of which in the network. The node is not trusted anymore when reputation score below the threshold $R_{thld}$.

**Definition 4: Term and Index.** Taking into account the feature of asynchronous in the distributed network, in order to ensure that process of consensus will be not affected by timestamp errors, the term and index which are numbered using consecutive integers to be used as logical timestamps. Only one leader exists per term, the current term number stored in each node. The index is used to uniquely identify the log that the leader node replicates to follower nodes to ensure that the order of logs in all nodes is consistent with the leader node.

### 4.2 Consensus Process Description

The PoR algorithm is divided into three steps: the leader and supervisor election phase, the reputation model computing phase, and the consensus phase. Nodes use Remote Procedure Call (RPC) to communicate in the network. The consensus process of PoR is shown in Fig. 5

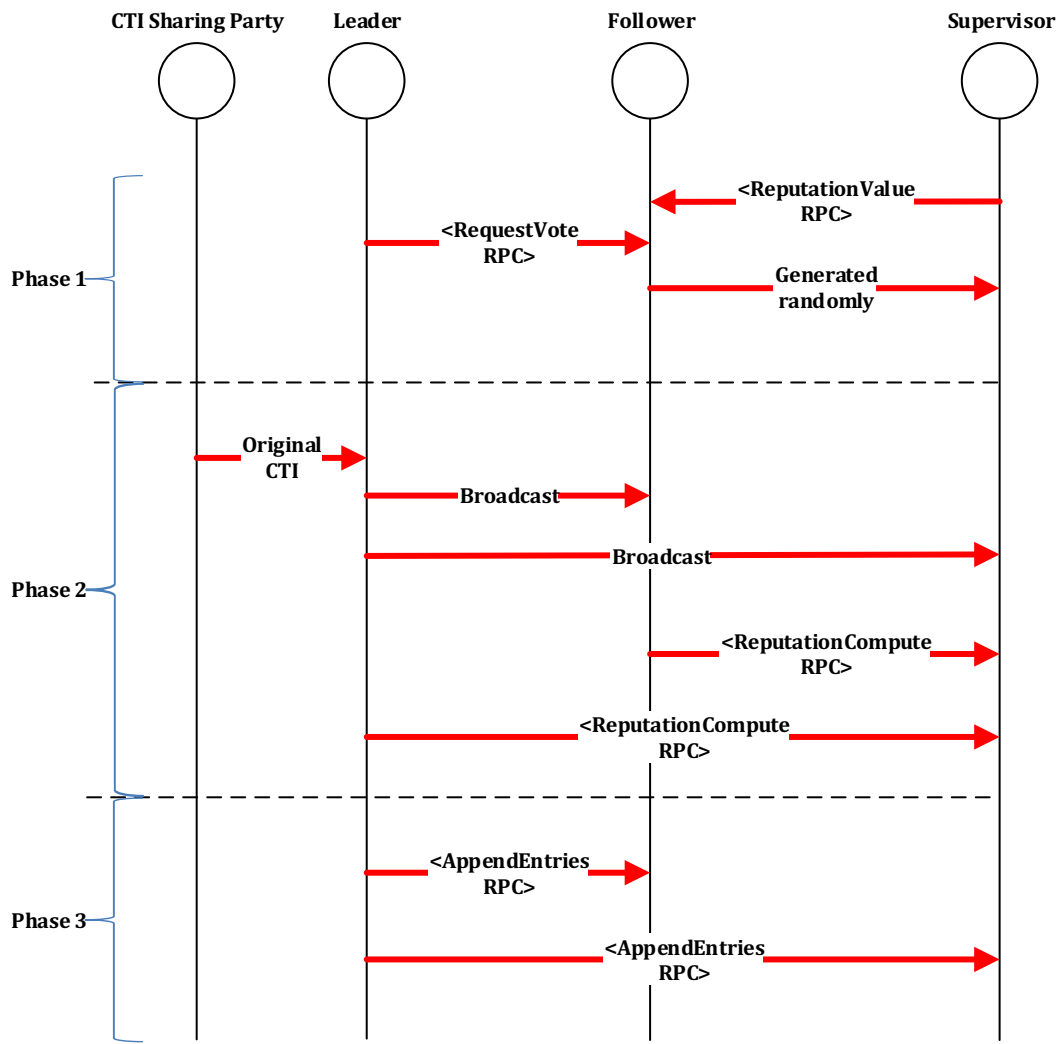

**Fig 5 The consensus process of PoR**

Phase 1 - Leader and Supervisor Election Phase.

All nodes are in a follower state when the consensus starts. If no heartbeat request that from the leader has been received for a period of time, a follower in the network becomes a candidate to start a new round of leader election and then sends out a vote request to all rest of the followers by RequestVote RPC. A node will only send a reply to the trusted candidate node that is certified by the supervisor node when receives a vote request, and the process is realized by ReputationValue RPC. The description of RequestVote RPC and ReputationValue RPC are shown in Table 3 and Table 4. The candidate will be elected as leader when received responses from the most follower nodes. The function of the leader is described in Definition 1. The leader will send a heartbeat request to all nodes in the network regularly to extend the term. The new term's supervisor node has generated randomly in the rest of the follower nodes based on the reputation score.

Table 3. The description of RequestVote RPC communication

| RequestVote RPC | |
| --- | --- |
| **Parameter** | **Description** |
| Term | The current term of the candidate node. |
| CandidateID | The id of the candidate node. |
| **Return** | **Description** |
| Term | The current term of this follower node. |
| VoteGranted | Set to true when the candidate won this vote. |

Table 4. The description of ReputationValue RPC communication

| ReputationValue RPC | |
| --- | --- |
| **Parameter** | **Description** |
| Term | The current term of the candidate node. |
| CandidateID | The id of the candidate node. |
| **Return** | **Description** |
| Term | The current term of this follower node. |
| TrustGranted | Set to true when the candidate is the faithful node. |

Phase 2: Reputation model Computing Phase.

Leaders transform threat intelligence that from CTI sharing parties into proposals that including the detailed information of CTI such as attacker's IP, attack characteristics, attack methods and so on, then broadcast to all follower nodes and supervisor node. The function of the proposal is in two aspects, hide the private information of the attacked party, and ensure that the consensus result is deterministic in a decentralized system by adding information such as term and index. To prevent the false positives of the proposal are generated by the leader, and increase the accuracy of CTI, the supervisor uses the approach of probabilistic that based on the detection result of all followers to determine the validity of the proposal's detailed information and calculate the reputation value by communicating with all followers in ReputationCompute RPC after received a proposal from the leader, which called reputation model. The description of ReputationCompute RPC is shown in Table 5, the computation of the reputation model will be elaborated in chapter 4.3.

Table 5. The description of ReputationCompute RPC communication

| ReputationCompute RPC | |
| --- | --- |
| **Parameter** | **Description** |
| Term | The current term of leader node. |
| NodeID | The id of the node. |
| PrevIndex | The index of consensus proposal immediately preceding new ones. |
| Entries[ ] | The detailed information of CTI detected by the node such as attacker's IP, attack characteristics or attack methods. |
| **Return** | **Description** |
| Term | The current term of leader node. |
| Success | Set to true when the detailed information of CTI detected by the node is valid. |

Phase 3: Consensus Phase.

Store the transaction of CTI proposal that into a block is a permitted operation when the most members in CTI sharing collaboration consortium agree with it. The leader node broadcasts an AppendEntries RPC that including the detailed information of CTI proposals judged to be valid to all follower nodes when the responses of the reputation model from the supervisor have been received. The description of AppendEntries RPC is shown in Table 6.

Table 6. The description of AppendEntries RPC communication

| AppendEntries RPC | |
| --- | --- |
| **Parameter** | **Description** |
| Term | The current term of leader node. |
| LeaderID | The id of leader node. |
| PrevIndex | The index of consensus proposal immediately preceding new ones. |
| Entries[ ] | Proposal entries to store in each follower node (empty for heartbeat request). |
| LeaderCommit | The commitIndex of leader node. |
| **Return** | **Description** |
| Term | The current term of leader node. |
| Success | Set to true when verification of proposal that from leader is passed. |

Each follower node that receives the AppendEntries RPC confirms the correctness of the proposal in that message to the leader when verification is passed, the standard of a correct proposal as shown in Table 7. A consensus has been reached when the leader receives verification responses from the supervisor and more than 51% of followers in the network. Then each follower node records the information of the proposal along with the term and index number on their local blockchain.

Table 7. The description of correct proposal in consensus phase

| Index | Criteria |
|---|---|
| Term | Leader's term >= Follower's term. |
| PrevIndex | The prevIndex of this proposal's is more than the immediately preceding new ones. |
| Entries[ ] | This proposal's detailed information that from leader is same as the responses of reputation model from supervisor. |

### 4.3 Reputation model

Naive Bayes algorithms as an instance to demonstrate the Reputation model that proposed in this paper. Let eigenvector $X = \{x_1, x_2, \dots, x_k\}$ indicates to the detailed information that detected by follower node $\{n_1, n_2, \dots, n_k\}$ from received proposal, where $k$ is number of follower nodes that provided detailed information of proposal. An example of $x_i$ can be: $(IP, method, tool, characteristics, TTPs, \dots)$, which indicates the detailed information of attack behavior such as attacker's IP address/domain, attack tool/method, HASH, MD5, static/dynamic behavior characteristics and other attributes detected from CTI propose that need to share with peer partner.

There are $N$ nodes in the network. The proportion of follower nodes that submit detailed information to supervisor node by ReputationCompute RPC in phase 2 is $P(y) = k/N$. The probability of a valid CTI proposal that determined by follower nodes can be written as $P(y \mid X)$. Assume that the node provides information independently, then the equation can be further written as follows by using Bayes' Theorem.

$$P(y|X) = \frac{P(y)*P(X|y)}{P(X)} = \frac{P(y)*\prod_{i=1}^{i=|k|}P(x_i|y)}{P(X)} \quad (1)$$

The consensus is reached among the CTI sharing collaboration consortium by evaluating the credibility of proposal the received from the leader node in the method that check eigenvector $X$. Supervisor node calculates the reputation score of leader node and each follower node based on the credibility of information eigenvector. We only pick the information eigenvector of proposal with $P(y|X) \geq P(T)$, which means *correct detection*, where $P(T)$ is the threshold that set by situation among CTI sharing collaboration consortium. Equation (2) and (3) is a calculation method of correct detection or incorrect detection.

$$correct\ detection: P(y|X) \geq P(T) \quad (2)$$

$$incorrect\ detection: P(y|X) < P(T) \quad (3)$$

Reputation score of node i that submitted detailed information of correct detection ($x_i \in X$) will be increase. Reputation score of node i that not submitted detailed information of correct detection ($x_i \ not\ in\ X$) will be decrease. The method for calculation of the reputation score in node i can be expressed as equation (4) and (5), where $M$ indicates the reputation weight that used for further incentives or penalties.

$$R_i = R_i + M * \frac{\sum_{t_{0,i}}^{t_{current,i}} correct\ detections}{\sum_{t_{0,i}}^{t_{current,i}} proposals} \quad (4)$$

$$R_i = R_i - M * \frac{\sum_{t_{0,i}}^{t_{current,i}} incorrect\ detections}{\sum_{t_{0,i}}^{t_{current,i}} detection} \quad (5)$$

As defined in definition 3, when a new node joins the CTI sharing system, its reputation score is $R_{init}$, if the node matins normal behavior in proposal detected, its reputation score $R_i$ will increase and have more opportunities to be leader node or supervisor node. The node is unfaithful whose reputation score $R_i$ is lower than the threshold $R_{thld}$. If leader node is unfaithful node, which means its qualifications will be terminated in this term. Reputation model proposed in this paper can reduce the impact of unfaithful behaviors that under malicious attacks or wrong decision.

## 5 PERFORMANCE AND EVALUATIONS
### 5.1 Performance of the PoR algorithm

The PoR algorithm in consortium blockchain CTI sharing model that proposed can achieve byzantine fault tolerance and defense against blockchain attacks such as Sybil attacks. Compared to main consensus algorithm of consortium blockchain, we make an analysis of the security and performance of PoR algorithm and summarize them in Table 8.

Table 8. Performance in the consensus algorithm of consortium blockchain

| Algorithms | PBFT | Raft | PoR |
|---|---|---|---|
| Crash Fault Tolerance | 33% | 50% | 50% |
| Byzantine Fault Tolerance | 33% | N/A | 50% |
| Time complexity | $O(n^2)$ | $O(n)$ | $O(n)$ |
| Security | Strong | Weak | Strong |

Crash Fault Tolerance represented the fail-stop or crash failure that not malicious behaviors happened in blockchain system. Byzantine Fault Tolerance represented the byzantine behaviors happened in blockchain system such as tamper or submit wrong information.

**Hypothesis.** PoR consensus algorithm can achieve byzantine fault tolerance. The detailed information of CTI proposal can be detected and shared correctly by PoR algorithm when the number of byzantine nodes is less than 1/2 of all nodes.

**Proof.** The byzantine fault tolerance of PoR depends on reputation model in consensus algorithm. Naive bayes algorithms as an example of reputation model in this paper, assume that the number of byzantine nodes in the network is f, supervisor node in PoR can analyze correctly byzantine behaviors from eigenvector $X$ composed of detailed information that detected by all follower nodes when the total number of nodes in CTI sharing collaboration consortium network more than 2f+1. So PoR algorithm can tolerate at most 50% byzantine nodes or crash nodes.

Time complexity represented the communication cost and scalability of consensus algorithm. Adding blocks to blockchain in PBFT needs verification by communicated in each two nodes and three-phase commit, the time complexity of PBFT is $O(n^2)$. Consensus processes in Raft and PoR only require the leader nodes to send messages to the follower nodes, there is no need to communicate between the follower nodes. So, the time complexity of Raft and PoR is $O(n)$.

**5.2 Evaluations**

We conducted the experiments using on a computer with an Intel Core i5 and 16 GB RAM running MacOS operating system. The construction of the reputation model is implemented using Python3.6, the PoR consensus algorithm is using Golang1.14.7. We created a test network in a simulation environment to confirm our approach can meet the requirement of CTI sharing and exchange.

We compare the PoR-based CTI sharing model with other consortium blockchain-based CTI sharing models that use different consensus algorithm discussed below.

(1) Byzantine Fault Tolerance Consensus Based Model. Store the proposal of CTI that into a block is a permitted operation when confirmed by most members in CTI sharing collaboration consortium. In order to prevent Byzantine attacks that happened in network, every two nodes need to verify with each other to confirm CTI proposal. The typical example of this model is Tendermint [33].

(2) Crash Fault Tolerance Consensus Based Model. Store the proposal of CTI that into a block is a permitted operation when the most members in CTI sharing collaboration consortium agree with it. This model can achieve low latency and high throughput, but it only be used in a non-byzantine environment. The typical example of this model is HyperledgerFabric [34] (v1.4 and above).

**Experiment 1: The security of the PoR-based CTI sharing model**

The security of the PoR-based CTI sharing model is measured by the cost of time in distinguish byzantine nodes in the network. We use the metrics of 'Quality of Detection in Byzantine Node (QoD)' to quantify the performance in security, the calculation method of QoD is described in equation (6), where $\sum Consensus$ means the time consume of total CTI proposal in reach consensus, $\sum ByzantineNode$ indicates the the time consume that all byzantine nodes were determined to be unfaitful node in the network.

$$QoD = \frac{\sum_{t_{0,i}}^{t_{current,i}} ByzantineNode}{\sum_{t_{0,i}}^{t_{current,i}} Consensus} \qquad (6)$$

The experiment has been simulated under various conditions that the reputation weight, the threshold of reputation score, the probability of byzantine behavior and the number of nodes. Experiment result in Fig 6 illustrated that the time increase as the proportion of the byzantine nodes and the scale of the sharing collaboration consortium varies.

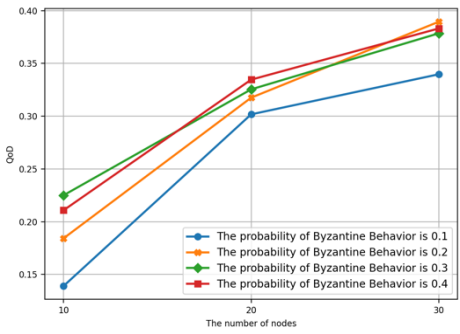
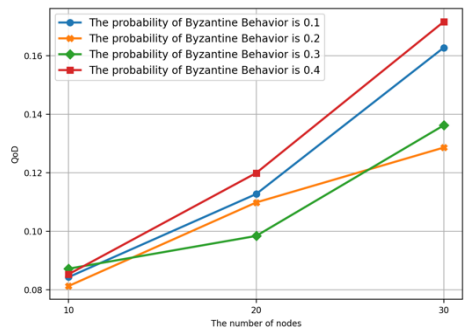

a. $R_{init}$=50, $R_{thld}$=10, $M$=5　　　　　　b. $R_{init}$=50, $R_{thld}$=10, $M$=15

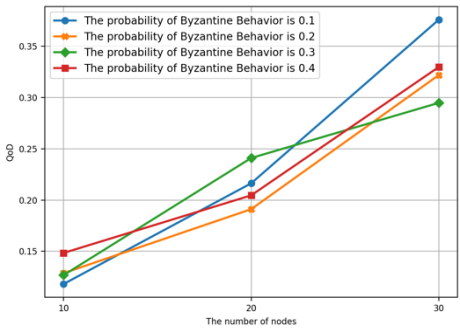
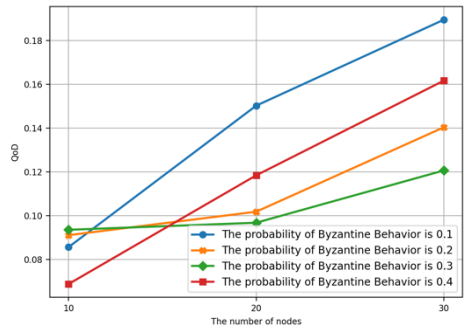

a. $R_{init}$=50, $R_{thld}$=20, $M$=5　　　　　　b. $R_{init}$=50, $R_{thld}$=20, $M$=15

**Fig 6 The security performance of PoR-based CTI sharing model**

**Experiment 2: The efficiency comparison of different sharing model**

The method of our evaluation is measuring the efficiency by latency and throughput. Latency refers to the time required for a single proposal of CTI to reach the consensus on the whole network, the process of a proposal update in the blockchain including reputation model computing phase and consensus phase. The experiment compares the latency between the PoR-based CTI sharing model and other blockchain CTI sharing models, as shown in Fig 7. Although the latency of our approach is worse than CFT-based model about 20% due to confirmation mechanism of reputation model in the PoR algorithm, it is still remarkably better than CFT-based model.

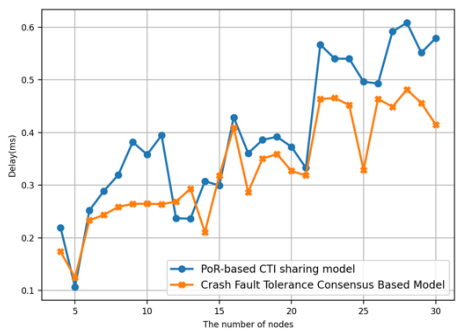
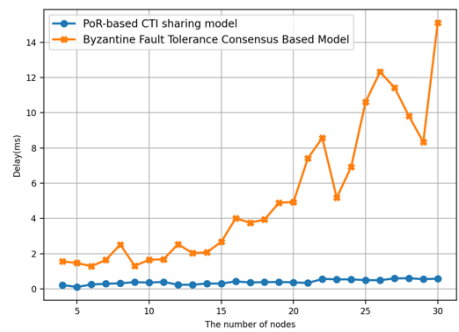

a. Compared with CFT-based model　　　　　b. Compared with BFT-based model

**Fig 7 Latency to reach consensus with different nodes**

Throughput is represented in PoR consensus algorithm as the number of transactions of CTI proposal that reach a consensus over time. We use ten client nodes to generate 1000 transactions of CTI proposal, and calculates the corresponding throughput based on the time required to reach a consensus under different numbers of transactions. As shown in Fig 8, with the number of node is further increased, the throughput of our proposed approach is better than BFT-based model. In addition, there have loss about 30% in throughput compared with CFT-based model because of byzantine fault tolerance supported by our approach.

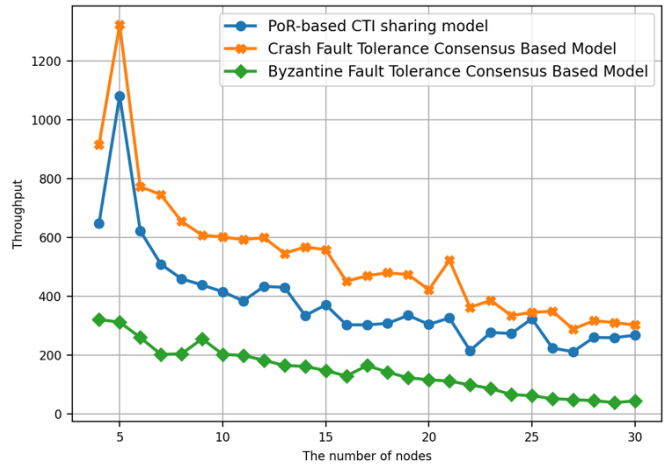

Fig 8 The performance of throughput with different nodes

**5.3 Summary**

In the simulation environment, compared to Crash Fault Tolerance Consensus Based Model, the PoR-based CTI sharing model requires an additional reputation computing process, so there is a loss in efficiency of consensus. However, compared to the Byzantine Fault Tolerance Consensus Based Model, our model still has advantages in latency and throughput performance. Thus, our results show that the PoR-based CTI sharing model reach better performance balance in speed, scalability, security, and byzantine fault tolerance.

# 6 CONCLUSIONS

This paper's contributions include a novel cyber threat intelligence (CTI) sharing approach using consortium blockchain that leverages advancements in consortium blockchain and distributed reputation management systems to automated process and defend against cyber-attack threats, as well as consensus algorithm called PoR (Proof-of-Reputation) that based reputation model to meet the effectiveness and security requirement. We devised three test scenarios in simulation environment to evaluate the proposed approach. Our evaluation results from simulation results show that proposed PoR-based CTI sharing model can achieve the needs of exchange of threat intelligence data in performance of speed, scalability and security, thus it can be applied to the scenarios of CTI sharing and exchange.